\documentclass[twocolumn,preprintnumbers,amsmath,amssymb,floatfix]{revtex4}% showpacs
\usepackage{epsfig,graphicx,dcolumn,bm}
\begin{document} 

\title{Relaxation in {\it open} one-dimensional systems}

\author{Prasanth P Jose \footnote[2]{jose@sscu.iisc.ernet.in}
and Biman Bagchi\footnote[3]{bbagchi@sscu.iisc.ernet.in}}

\affiliation{Solid State and Structural chemistry Unit,
Indian Institute of Science,
Bangalore - 560012, India.}
\date{\today}

\begin{abstract}
A new master equation to mimic the dynamics of a collection of
interacting random walkers in an {\it open system} is proposed and
solved numerically.
In this model, the random walkers interact through excluded
volume interaction ({\it single-file system}); and the total
number of walkers in the lattice can fluctuate because
of exchange with a bath.
In addition, the movement of the random walkers is biased  by an
external perturbation.
 Two models for the latter are considered: (1) an inverse potential 
($V \propto \frac{1}{r}$), where $r$ is the distance between 
the center of the perturbation and the random walker and (2)
 an inverse of sixth power potential ($V \propto \frac{1}{r^6}$).
The calculated density of the walkers and the total energy
show interesting dynamics.
When the size of the system is comparable to the range of 
the perturbing field, the energy relaxation is found to be
highly non-exponential.
In this range, the system can show stretched exponential 
($e^{-{(t/\tau_s)}^{\beta}}\;$) 
and even logarithmic time dependence 
of energy relaxation over a limited range of time. 
Introduction of density exchange in the lattice
{\it markedly weakens} this non-exponentiality of the
relaxation function, irrespective of the nature of perturbation.

\end{abstract}

\maketitle

\section{Introduction}
Relaxation dynamics of interacting particles in a 
 one-dimensional system is often difficult to understand 
because the traditional coarse-grained descriptions 
(such as hydrodynamics or time dependent mean-field 
type approximations) fail in this case.
This is because of the existence of long-range correlations
mediated through the excluded volume interaction.
In such cases, random walk models have often proved to be
successful in describing the non-exponential relaxation 
commonly observed in one-dimensional systems.

Recently, several experimental studies have reported 
energy relaxation in one-dimensional or quasi one-dimensional 
systems such as DNA\cite{berg,zew}.
These experiments employed the time dependent fluorescence Stokes
shift (TDFSS) technique to gather a quantitative measure of the
time scale involved.
In one of these experiments Bruns {\it et al.}\cite{berg} have
calculated structural relaxation of DNA oligonucleotides. 
They found that the red shift of the 
fluorescent spectrum follows a {\it logarithmic} time
dependence.
The experiment described above is an example where
dimensionality plays an important role in the energy 
and density relaxation. 

In this work, we propose a random walk model for the carriers 
in a one-dimensional channel to mimic the relaxation of
energy and density found in experiments described above. 
There were several theoretical studies devoted to random walk model
in one-dimensional systems\cite{c1,c2,c3,c4,hub1,rich1,rich2}.
Many recent studies based on random walk model in one-dimensional 
systems have explored transport and other dynamical properties such
as, DC conductivity, frequency dependent conductivity, effect
of bias on diffusion, space and time dependent probability 
distribution of random walkers, mean residence time and mean
first passage time of random walkers etc.
\cite{rich2,yu,pit,kl1,kl2,file1,file2,file3}.

The work presented here is based on a master 
equation for random walk in a {\it single-file} system 
(a one-dimensional system that does not allow particles
to pass through each other) that allows exchange of the
particles with the bath (see figure 1 for a schematic 
illustration).
\begin{figure}[htb] 
\epsfig{file=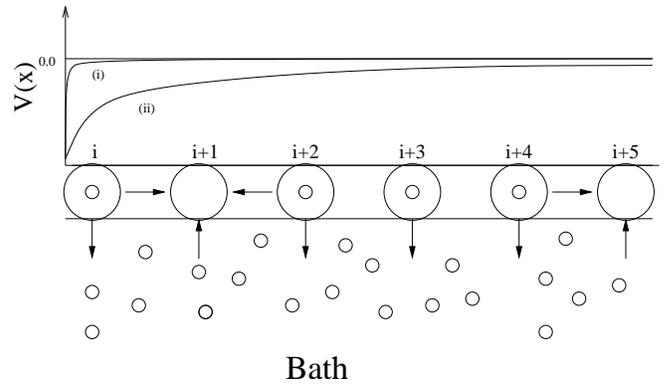,width=8.6cm} 
\caption{The perturbation potential and the
 various transitions are shown schematically. The attractive part 
of the  Lennard-Jones potential is shown by the line (i) 
and Coulomb potential is given by the line (ii).}
\end{figure} 
Here we employed two model potentials with different 
characteristics to study the effect of perturbation on the
density and energy relaxation processes. 
The potentials used here are (1) inverse of distance
potential (example is a Coulomb's field generated
by a trapped charge in the lattice) and (2) inverse of 
sixth power of distance potential (this is a short range
interaction compared to Coulomb interaction and is taken
as attractive part of the Lennard-Jones interaction). 
Energy and density relaxation functions in a one-dimensional 
channel without number fluctuation is calculated and then
compared with that of a channel, where
particle number fluctuates due to exchange with the bath.
It is observed that density fluctuation in the channel makes 
energy relaxation exponential.
 
The rest of this paper is organized as follows.   
Section 2 gives the  description of 
the master equation used in the simulation. 
The details of the Monte Carlo simulations are 
given in the section 3.
Results of  the simulations were analyzed in the section 4. 
Section 5 presents the conclusions from the Monte Carlo 
simulations. 
 
\section{The Generalized Master equation} 
 
Let the total number of lattice sites in a system consisting
of a linear lattice and a bath be $\mathcal{N}_T=\mathcal{N}_L
+\mathcal{N}_B$, where $\mathcal{N}_L $ is the number of lattice
sites of the lattice and  $\mathcal{N}_B$ is the number
of accessible bath sites.
The time dependent probability $P^L_i(t)$ of finding a
particle in the $i\;th$ site of the lattice at time t
is given by the following master equation 
\begin{eqnarray} 
\frac{dP^L_i(t)}{dt}&=&\sum_{m=1}^{\mathcal{N}_L\;\;\prime}
 w_{m,i}(t)P^L_m(t)-w_{i,m}(t)P^L_i(t)\nonumber\\ 
&&-P^L_i(t)\sum_{j=1}^{\mathcal{N}_B} w_{i,j}(t) 
+\sum_{j=1}^{\mathcal{N}_B}w_{j,i}(t)P^B_j(t), 
\end{eqnarray} 
(the prime on the summation signifies that the term $i=m$ 
is to be omitted from the sum) where $w_{m,i}(t)$ gives the
transition probability of the particle 
from site $m$ to $i$ per unit time.
The last two terms of the master equation introduce
density fluctuation in the lattice by allowing exchange of
particles with the bath.
In the master equation, the index $j$ sums over the bath
sites and ${P^B}_j$ is the probability of finding a particle 
in the $j\;th$ bath site. 
This master equation can be simplified by assuming 
that lattice sites in the bath are numerous ( $\mathcal{N}_B >>
\mathcal{N}_L $ ) and the  number of walkers in the bath
is much larger than that in the lattice.
These assumptions allow us to perform an averaging over the 
bath sites. 
Summation in the last two terms of the equation (2) can be omitted
by substituting the properties of the bath with 
that of an ideal bath defined below.
The sum of the transition probability $w_{i,j}(t)$ to the
bath sites can be replaced by the term $w_{out}(t)$ 
($\sum_{j=1}^{\mathcal{N}_B} w_{i,j}(t) = w_{out}(t)$), since
$w_{i,j}(t)$ depends only on the temperature, size and number
of particles in the system (discussed in detail later in this
section).
This ideal bath has infinite capacity to absorb particles.
Similarly, when a particle is absorbed by the lattice this bath 
site acts as a supplier of infinite number of particles. 
The transition probabilities from all lattice sites to the bath 
are equal and assumed to be independent of the instantaneous
state of the bath.
Therefore, $w_{j,i}$ can be replaced by $w_{in}$  
and $\sum_{j=1}^{\mathcal{N}_B}P_j^B$ can be replaced 
by $P_{bath}$.
The resultant master equation is simpler and is given by
\begin{eqnarray} 
\frac{dP^L_i(t)}{dt}&=&\sum_m^{\mathcal{N}_L\;\;\prime}
w_{m,i}(t)P^L_m(t) -w_{i,m}(t)P^L_i(t)\nonumber\\ 
&&-w_{out}P^L_i(t)+w_{in}(t)P_{bath}.
\end{eqnarray}
If only nearest neighbor exchanges are
allowed then the master equation can be written as  
\begin{eqnarray} 
\frac{dP^L_i(t)}{dt}&=& w_{i+1, i}(t)P^L_{i+1}(t)+
w_{i-1, i}(t)P^L_{i-1}(t)\nonumber\\
&&-w_{i, i+1}(t)P^L_i(t)-w_{i, i-1}(t)P^L_i(t) \nonumber\\ 
 &&-w_{out}(t)P^L_i(t) + w_{in}(t)P_{bath}.
\end{eqnarray} 
At equilibrium this system represents a grand canonical ensemble.
Hence for insertion and removal of the particles in the randomly 
selected sites we use grand-canonical Monte Carlo method, 
where probability ($\mathbf{P}_{\mu,L,T}$) of system having
number of particles $N$ is\cite{fr} 
\begin{equation} 
\mathbf{P}_{\mu,L,T} \propto \frac{L^N}{\Lambda^{N} N!}\; 
e^{-\beta[\; E-\mu N]}, 
\end{equation} 
where $\beta$ is the inverse of Boltzmann constant
times absolute temperature ($\frac{1}{k_BT}$), 
$\Lambda$ ($=\sqrt{\frac{h^2}{2\pi m k_B T}}$) is the thermal 
de-Broglie wavelength, $L$ is the length of the linear lattice,
$\mu$ is chemical potential and $E$ is 
the potential energy of the linear lattice.
Particles of this system obey Boltzmann distribution at 
equilibrium, hence the transition probability $w_{i,i+1}(t)$
can be calculated from the total energy cost for the move.
The  hopping probability is calculated for a particle 
from one site to another using Metropolis scheme \cite{fr}.   
Hopping  probability of a random walker from one site to the 
neighboring site of the lattice is given by
\begin{equation} 
w_{i,i+1}=min[1,\;e^{-\beta \Delta E}].  
\end{equation} 
 The transition probabilities for absorption  and
 desorption from the lattice  can be obtained 
\cite{fr} so as to satisfy the detailed balance at equilibrium as 
\begin{eqnarray} 
w_{in}&=&min[1,\;\frac{Q}{{(N+1)}}\;e^{-\beta\;\Delta E}],\\ 
w_{out}&=&min[1,\;\frac{N}{Q}\;e^{\beta\;\Delta E}], 
\end{eqnarray} 
where $Q$ is  given by $\frac{L}{\Lambda}\;e^{\beta \mu} $.
A non-trivial aspect of this master equation is that
the transition probability $w_{in}(t)$
{\it varies with time} in response to the  number fluctuations.
This explicit time dependence of $w_{in}(t)$ pose 
formidable  difficulty in obtaining an explicit analytic 
solution of the problem.
In this system of interacting particles, the transition
probability to neighboring sites depends on the
instantaneous probability of that site being occupied.
In a one-dimensional channel with hard rod interactions  
between the carriers, mobility of each carrier is restricted to  
a portion of the lattice. 
The particles move in the linear lattice under the influence
of a chosen  potential, which is  at a fixed position  
of this lattice.
Hence the energy cost for hopping in the lattice is 
\begin{displaymath} 
\Delta E= \left\{ \begin{array}{ll} 
E_1
& \mbox{(transition to vacant site)}\\ 
\infty & \mbox{(transition to occupied site)} 
\end{array} 
\right.  
\end{displaymath}
where
$E_1=K_1( \frac{1}{ x_i} - \frac{1}{x_{i+1}})$ for
the  potential 1 and 
$E_1=K_2( \frac{1}{ {x_i}^6} - \frac{1}{{x_{i+1}}^6})$
for the potential 2.
For  absorption of particles to the lattice at $i$th site 
\begin{displaymath} 
\Delta E=\left\{ \begin{array}{ll} 
E_2
& \mbox{(creation of vacant site)}\\ 
\infty & \mbox{(creation  of occupied site)} 
\end{array} 
\right. 
\end{displaymath} 
where 
$E_2=K_1 \frac{1}{x_i}$ for potential 1
and $E_2=K_2 \frac{1}{{x_i}^6} $ for potential 2.
 For the desorption of particles from the $i$th site 
\begin{equation} 
\Delta E=E_2. 
\end{equation} 
$Q$ in this system is constant and equal to the
initial number of the particles in the lattice. 
This assumption specifies the value of the chemical potential
of this linear lattice. 
 
\section{The details of  Monte Carlo simulation} 
 
The Monte Carlo simulations are carried out on a linear 
lattice, with 50\% (on the average) of the lattice sites are
occupied by the random walkers. 
Closed boundary condition is used to explore the size 
effect on the relaxation. 
In this one-dimensional lattice, the adjacent  sites are equally
spaced and all the sites are identical in energy in the absence
of the perturbation. 
Here inhomogeneity in the lattice is generated by 
the perturbation of potential introduced in the lattice. 
In addition, there is a dynamic  disorder in the
lattice that  originates  from the instantaneous
rearrangement of the interacting random walkers.
The initial configuration of the random walkers 
in an unperturbed lattice is chosen from a random
distribution, such that no two particles occupy the same site.
The inverse potential  arise from Coulomb interaction, 
hence the magnitude of the biasing potential is calculated from
the interaction between the  charge of the carriers ($q_1$) and 
the charge at the center of bias ($q_2$) in a medium of 
dielectric constant $\epsilon$. 
The constant $K_1$ for potential energy can be calculated  as
$K_1=\frac{q_1 q_2}{\epsilon}$.
Here both the perturbing charge and the charge 
of the carrier have magnitude of one electron 
and their signs are opposite.
Assuming high screening effect, the value of $\epsilon$ 
is  taken as equal to that of water, that is  80.
The distance between the adjacent sites in the linear
lattice is 4\AA (approximately the vertical distance
between two DNA base pairs in the double helix).
All simulations were conducted near room temperature (300K).
 The value of $K_1$ used in the simulation is -1.7$k_B T$.
Compared to Coulomb's interaction, the Lennard-Jones interaction is
short ranged, hence the constant $K_2$ is kept high to extend
the range of the potential, value of $K_2$ used is -10 $k_B T$. 

The non-equilibrium Monte Carlo simulation starts from a
randomly chosen initial configuration.
Then a perturbing potential is introduced in the lattice at
at time $t=0$.
Subsequent relaxation of energy is monitored.
The simulation is repeated with different initial configurations
and the results are averaged.
In the Monte Carlo simulation movements are chosen randomly
such that no two events can occur at the same time. 
In this simulation one Monte Carlo step is equivalent to one
unit of time. 
Total potential energy of the lattice at any instant of time
when perturbed by Coulomb potential is given as
\begin{eqnarray} 
E(t)=\sum_i \kappa_1 \frac{1}{x_i} P_i(t), 
\end{eqnarray} 
where $\kappa_1$ is a constant.
Similarly, for the simulations that use  attractive Lennard-Jones 
potential  as the perturbation, the instantaneous potential energy 
of the lattice is given by
\begin{eqnarray} 
E(t)=\sum_i \kappa_2 \frac{1}{x_i^6} P_i(t), 
\end{eqnarray} 
where $\kappa_2$ is a constant.

For comparison of relaxation function for different system
sizes, we calculate dimensionless energy relaxation function
$S(t)$ which can be defined as 
\begin{equation} 
S(t)=\frac{E(t)-E(\infty)}{E(0)-E(\infty)}, 
\end{equation} 
where $E(t)$ is the instantaneous energy at time 
$t$ and $E(\infty)$ is the average energy of the lattice,
in equilibrium with perturbation.
Density relaxation is measured in terms of a
dimensionless quantity 
\begin{equation} 
C(t)=\frac{X(t)-X(\infty)}{X(0)-X(\infty)}, 
\end{equation} 
where $X(t)$ is defined as $X(t)= \sum_i x_i P_i(t)$.
 
In the equilibrium simulation the system is allowed to 
equilibrate with perturbation for 10,000 steps to get the 
initial configuration. 
The fluctuations in total energy of the system can be
defined as $F(t)\;=\;<E(0)E(t)>$.
Here a dimensionless energy fluctuation relaxation 
function can be defined as 
\begin{equation} 
S(t)=\frac{F(t)-F(\infty)}{F(0)-F(\infty)}.
\end{equation} 
 
\section{Results and Discussions} 

\subsection{Relaxation under Coulomb potential} 

\begin{figure}[htb] 
\epsfig{file=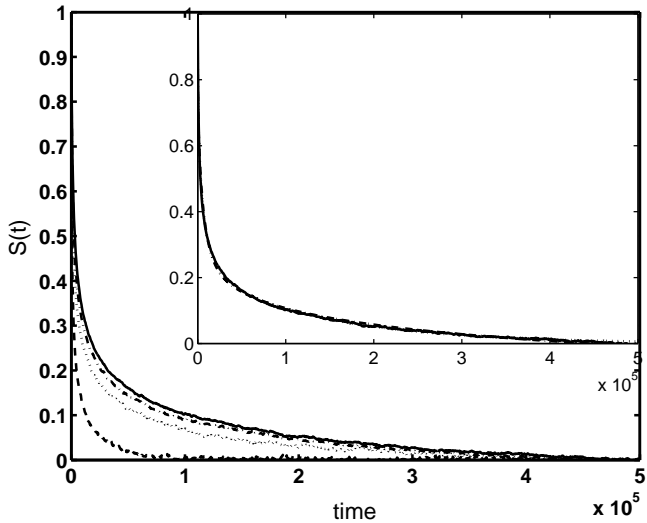,width=8.6cm} 
\caption{Energy relaxation function S(t) is plotted for four different
system sizes under Coulomb potential: $\mathcal{N}_L$ =50 
(dashed line), $\mathcal{N}_L$=100 (dotted line),
 $\mathcal{N}_L$=150 (dash-dot line) 
and $\mathcal{N}_L$=200 (continuous line)in a closed system.
 The inset shows the fit of the energy relaxation function for 
$\mathcal{N}_L$=200 using function $f_1(t)$ (dotted line)
 and $f_2(t)$ (dash-dot line).}
\end{figure} 
\begin{figure}[htb] 
\epsfig{file=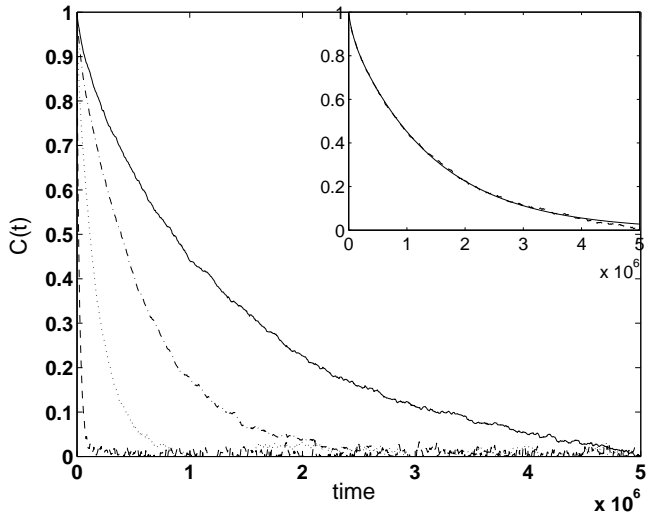,width=8.6cm} 
\caption{The density relaxation function $C(t)$ is plotted in a closed
system for four system sizes under Coulomb  potential:
 $\mathcal{N}_L$=50 (dashed line), $\mathcal{N}_L$=100 (dotted line),
$\mathcal{N}_L$=150 (dash-dot line) and $\mathcal{N}_L$=200
 (continuous line). The inset shows double exponential fit
 (continuous line) for $C(t)$ at $\mathcal{N}_L$=200.}
\end{figure} 
\begin{figure}[htb] 
\epsfig{file=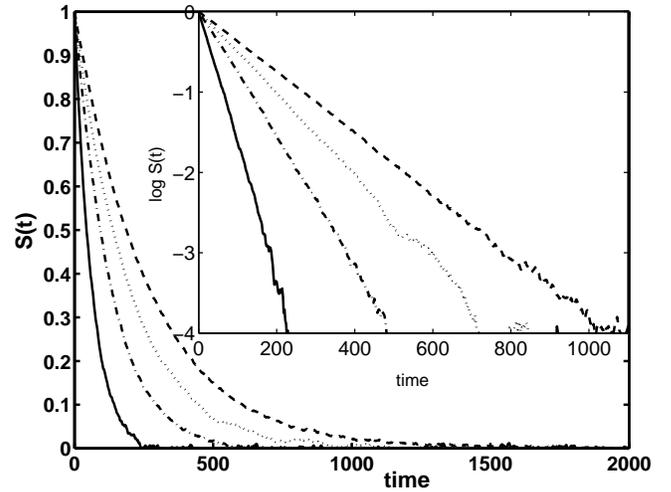,width=8.6cm} 
\caption{Energy relaxation function is plotted for different system 
lengths ($\mathcal{N}_L$=50 (continuous line),
$\mathcal{N}_L$=100 (dash-dot line), $\mathcal{N}_L$=150
(dotted line) and $\mathcal{N}_L$=200 (dashed line)) 
under the perturbation of Coulomb potential in an open
system. The log of $S(t)$ versus t plot in the inset shows 
straight lines due to the exponential relaxation.}
\end{figure} 
\begin{figure}[htb] 
\epsfig{file=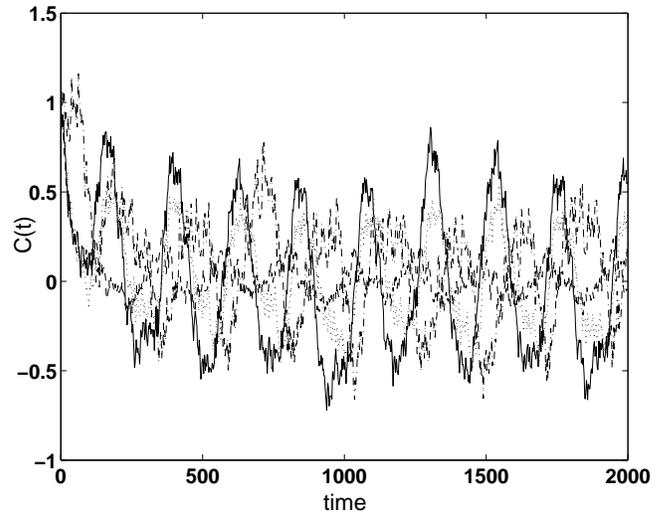,width=8.6cm} 
\caption{Density relaxation function with density fluctuation $C(t)$ 
is plotted here for four system sizes ($\mathcal{N}_L$=50 
(continuous line), $\mathcal{N}_L$=100 (dash-dot line),
$\mathcal{N}_L$=150 (dotted line) and $\mathcal{N}_L$=200
(dashed line)) under Coulomb potential. }
\end{figure} 
\begin{figure}[htb] 
\epsfig{file=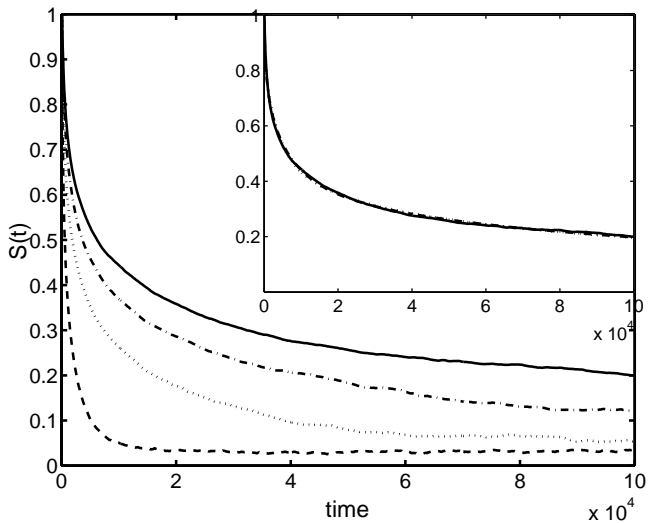,width=8.6cm} 
\caption{Energy fluctuation relaxation function is plotted for
a closed system at four sizes ($\mathcal{N}_L$=50 (continuous line), 
$\mathcal{N}_L$=100(dash-dot line), $\mathcal{N}_L$=150
(dotted line) and $\mathcal{N}_L$=200 (dashed line)) in
equilibrium under the perturbation of Coulomb potential.  
The inset shows the fit of the energy relaxation function
for $\mathcal{N}_L$=200 using function $f_1(t)$ (dotted
line) and $f_2(t)$ (dash-dot line). }
\end{figure} 
\begin{figure}[htb] 
\epsfig{file=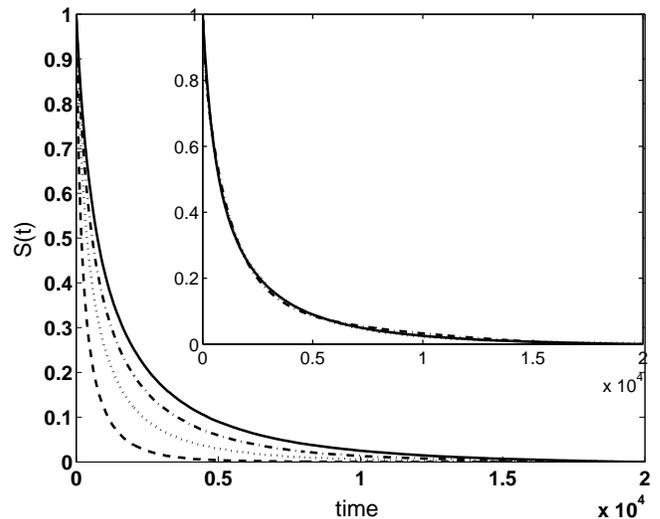,width=8.6cm} 
\caption{Dimensionless energy relaxation function $S(t)$ of a 
one-dimensional  channel with constant number density 
at different system sizes ($\mathcal{N}_L$=50 (continuous
line), $\mathcal{N}_L$=100 (dash-dot line), 
$\mathcal{N}_L$=150 (dotted line) and $\mathcal{N}_L$=200
(dashed line)) under the perturbation of Lennard-Jones
potential is shown here. The inset shows the fit of
the energy relaxation function for $\mathcal{N}_L$=200
using function $f_1(t)$ (dotted line) and $f_2(t)$ 
(dash-dot line). }
\end{figure} 
\begin{figure}[htb] 
\epsfig{file=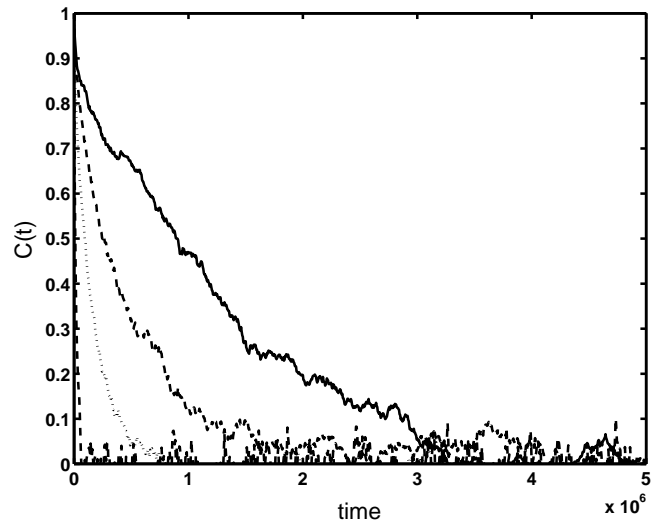,width=8.6cm} 
\caption{Density relaxation function  of the lattice under
the perturbation of Lennard-Jones potential at different
system sizes ($\mathcal{N}_L$=50 (continuous line), 
$\mathcal{N}_L$=100 (dash-dot line), $\mathcal{N}_L$=150
(dotted line) and $\mathcal{N}_L$=200 (dashed line)) for a
closed system is shown in the figure.}
\end{figure} 
\begin{figure}[htb] 
\epsfig{file=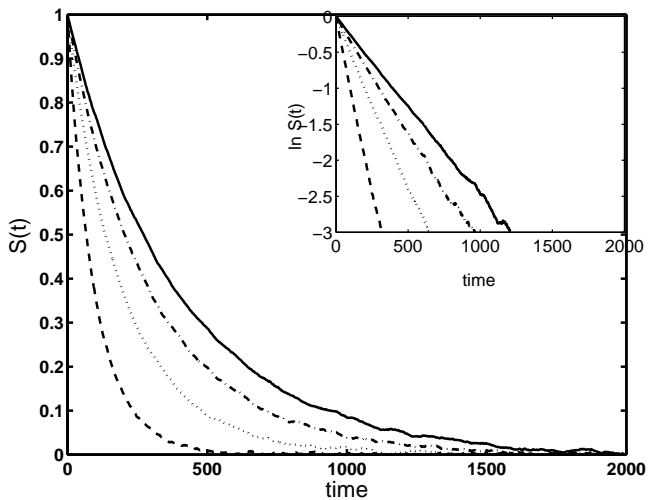,width=8.6cm} 
\caption{Energy relaxation function ($S(t)$) is plotted
at different system sizes ($\mathcal{N}_L$=50 
(continuous line), $\mathcal{N}_L$=100 (dash-dot line),
$\mathcal{N}_L$=150 (dotted line) and $\mathcal{N}_L$=200
(dashed line))with particle number fluctuations under 
Lennard-Jones potential. The log of $S(t)$ versus t plot
in the  inset shows straight lines due to the exponential
relaxation. }
\end{figure} 
\begin{figure}[htb] 
\epsfig{file=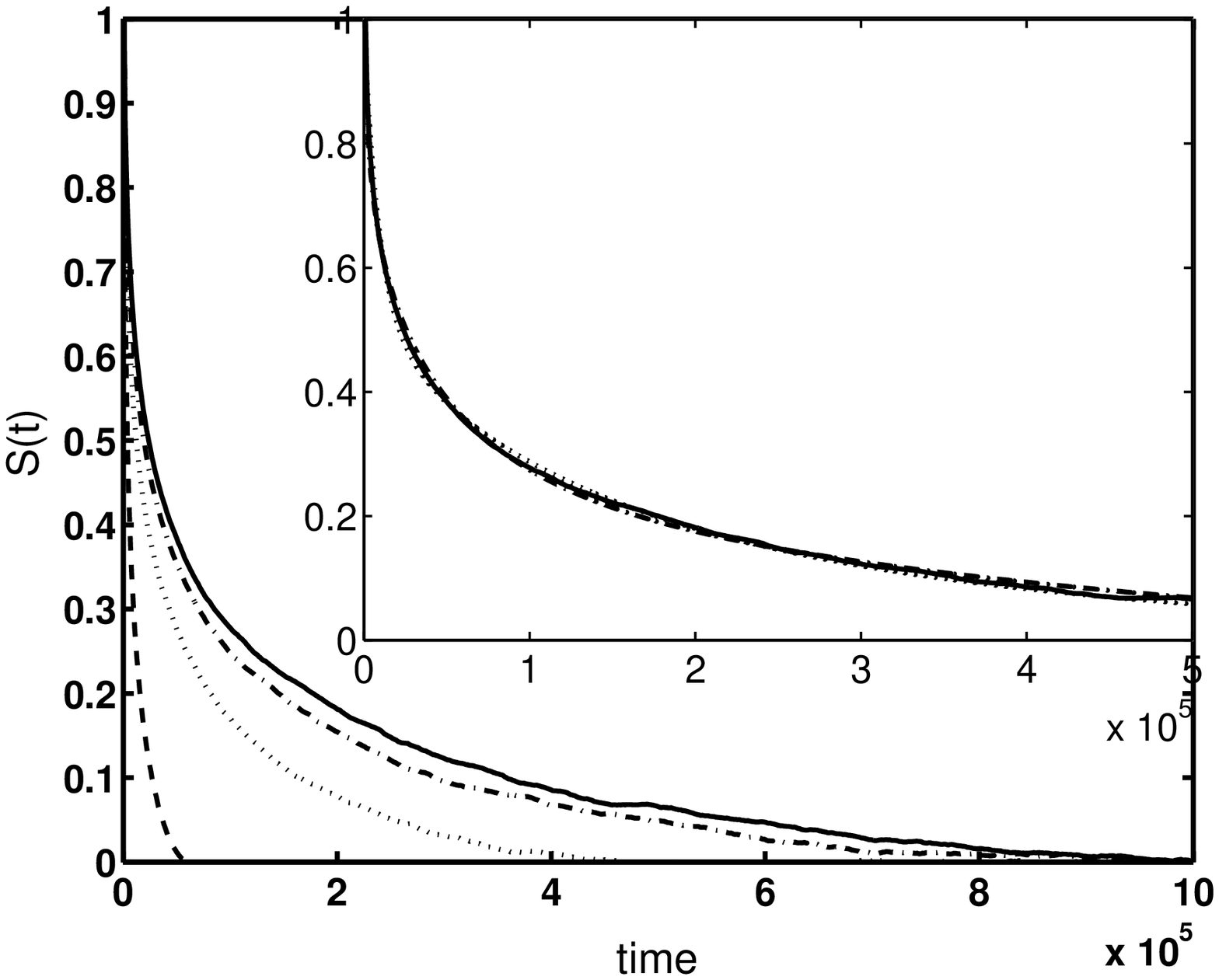,width=8.6cm} 
\caption{Energy fluctuation relaxation function is plotted for 
system in equilibrium under the perturbation of  
Lennard-Jones potential for different system sizes 
($\mathcal{N}_L$=50 (continuous line), $\mathcal{N}_L$=100
(dash-dot line), $\mathcal{N}_L$=150 (dotted line)
and $\mathcal{N}_L$=200 (dashed line)). The inset shows
the fit of the energy relaxation function for
$\mathcal{N}_L$=200 using function $f_1(t)$ (dotted line)
and $f_2(t)$ (dash-dot line)}
\end{figure} 
Figure 2 shows the energy relaxation function 
for different system sizes obtained from the non-equilibrium 
simulations of a one-dimensional channel without particle
fluctuation.
At short times, the potential energy of the lattice
 relaxes faster, as a response to the newly created
center of  perturbation.
This results in the accumulation of random walkers
near the center of the biasing field, which slows down
the relaxation rate.
At the same time, channels far from the center of 
perturbation remain active due to the reduction in
the density of particles in that region.
These effects together contribute to  the observed 
slowing down of the relaxation process after the  
fast initial decay. 
As  the system size increases the lowest possible energy
accessible to  the system obviously become lower, 
hence the relaxation becomes progressively slow as the  system 
size increases.
However, this effect becomes insignificant beyond
a limiting size of the lattice beyond which the strength of 
perturbation becomes negligible.
To analyze the  behavior of relaxation functions in 
the system, we have fitted the 
relaxation function to a sum of an exponential and a stretched 
exponential of the form
\begin{equation} 
f_1(t)=b_1e^{-t/\tau_1}+b_2e^{-(t/\tau_2)^{\beta}} 
\end{equation} 
(with constraints $b_1+b_2=1$ and $0\le b_1,b_2\le 1$).
In order to compare with the nature of relaxation with the 
results obtained by Bruns {\it et al.}\cite{berg},
a logarithmic function of the form 
\begin{equation} 
f_2(t)=1-c_1\;log_{10}(at)+c_2e^{-t/\gamma}
\end{equation}
is also used to fit the relaxation function. 
The inset of figure 2 shows the fit of the energy
relaxation function with $f_1(t)$ and $f_2(t)$, 
when size of the system is $\mathcal{N}_L$=200. 
Fitting parameters obtained for the function $f_1(t)$ 
 are $b_1=0.20,\;\tau_1=1.4\times10^5,\;
\tau_2=3.4\times10^5,\;\beta=0.52$.
 The fitting parameters for the function $f_2(t)$ are 
$c_1=0.16$ $a=2.5$ $c_2=0.25$ and $\gamma=7.1\times10^3$.
It is evident from the figure that the stretched 
exponential function and logarithmic functions both give
good description of  the energy  relaxation function 
in a one dimensional lattice without density fluctuations,
in the presence of the Coulomb potential.
The time dependence of  stretched exponential  
 \cite{kohl,ww} relaxation, is given by the
function of the form $S(t)=S_{0}\;e^{-(t/\tau_s)^{\beta}}\;$,
where $0<\beta<1$.
Theoretical explanation of the origin of  stretched 
exponential relaxation in the condensed matter have been addressed 
by many \cite{got1,got2,vlad,coh}, including Huber and coworkers
in a series of papers \cite{hub2,hub3}.
Their model is based on the following simple  master equation
approach 
\begin{equation} 
\frac{dP_i(t)}{dt}=\sum_{m\neq i} w_{m,i}P_m(t)-w_{i,m} 
P_i(t) 
\end{equation} 
where $w_{m,i}$ gives the transition probability of the particle 
from site $m$ to $i$. 
Stretched exponential relaxation can arise  
when there is a continuum of relaxation channels and the 
probability of any single channel being open is much less than unity.

Figure 3 shows the density relaxation function (which 
is a measure of particle aggregation) for different 
sizes of a system without number density fluctuations.
At $\mathcal{N}_L$=200,  a double exponential 
\begin{equation} 
f_3(t)=c_1\; e^{-t/\gamma_1}+c_2\;e^{-t/\gamma_2}
\end{equation}
(with constraints $c_1+c_2=1$ and $0 \le c_1,c_2 \le 1 $)
(shown in the inset of figure 3) fit of the density 
relaxation function reveals the presence of two distinct
time scales in the density relaxation.
 The fitting parameters obtained for the double exponential
($f_3(t)$) fit of density relaxation function are 
$c_1=0.90,\;\gamma_1=1.4\times10^{6},\; \gamma_2=8.3\times10^4$.

We now turn to the case when the number of walkers
in the lattice can fluctuate due to the exchange with the bath.
Figure 4 gives the energy relaxation function plotted for
different system sizes for this case. 
Note that, in this case, the energy relaxation is faster than 
the non-fluctuating case.
The exchange of particles with bath is equivalent to 
opening up of many wider channels of relaxation that dominate 
the relaxation process.
The log $S(t)$ versus $t$ plot in the inset of figure 4 
show straight lines, which is an evidence of 
exponential relaxation.
In this case walkers can bypass the obstacle caused by hard
rod interaction in  the path by exchange of particles with
the bath.
Figure 5 shows the density relaxation function of an  
open system for different sizes.
As system size increases, the effect of Coulomb field 
on the distribution of carriers decreases.
Hence the carrier density oscillates around a mean value
due to the density fluctuations and the effect
of perturbation in the density relaxation remains 
feeble and short lived.

The  energy fluctuation relaxation function for different sizes
of a system with constant number density and that is in equilibrium
with perturbation is shown in the figure 6. 
This relaxation function shows very slow non-exponential decay.
 Inset of figure 6 shows the energy relaxation function
obtained is  well fitted  by  $f_1(t)$ and
 $f_2(t)$. 
The fitting parameters obtained  for the $f_1(t)$ are 
$b_1=0.2,\;\tau_1=3.7\times10^3,\;\tau_2=3.6\times10^4,\;\beta=0.34$
and the fitting  parameters obtained for $f_2(t)$ are
 $c_1=0.21$ $a=0.06$ $c_2=0.13$ and $\gamma=6.9\times10^3$. 
Note that the energy relaxation function in non-equilibrium and
energy fluctuation relaxation function in equilibrium
show logarithmic time dependence.

\subsection{Relaxation under Lennard-Jones potential} 

The pronounced non-exponential decay found in non-equilibrium 
energy relaxation function for  Coulomb potential  
is also found in the case of attractive Lennard-Jones potential.
Figure 7 shows the energy relaxation function at different sizes for
a system without number fluctuation perturbed by the attractive
Lennard-Jones potential.
The fit of the energy relaxation function with $f_1(t)$ and 
$f_2(t)$ are shown in the inset of the figure 7.
The fitting parameters obtained for $f_1(t)$ are $b_1 \simeq 0,
\;b_2 \simeq 1.0,\; \tau_s= 1.1\times10^3,\; \beta = 0.62$
and $f_2(t)$ are $c_1=0.14$ $a=5.6\times10^1$ $c_2=0.58$
and $\gamma=1.3\times10^3$.
It is evident from the figure that, in the case of short-ranged 
interaction, the appropriate function which can fit relaxation
function is stretched exponential.
Here due to the large strength of the potential near the
perturbation center, the initial relaxation is driven and faster.
Corresponding density relaxation function for different sizes 
of this system is shown in the figure 8.

The energy relaxation function ($S(t)$ versus $t$) of a linear
lattice with particle number fluctuation is plotted for different
system sizes in the figure 9. 
The inset of the figure 9 shows log of $S(t)$ versus $t$ plot of 
the energy relaxation function, which show straight lines
that is a signature of exponential relaxation.
It is clear from the figures 4 and 9 that the number fluctuation
in the system makes the energy relaxation nearly exponential
irrespective of the nature and range of the perturbing potential.
 
The equilibrium energy fluctuation relaxation function  of
a system under the Lennard-Jones perturbation,  
for different system sizes, is plotted in the figure 10.
The energy fluctuation in this system is mostly from the
region where  potential is weak.
 The fit of the energy fluctuation relaxation with functions
 $f_1(t)$ and $f_2(t)$ are shown in the inset of the figure 10. 
The  fitting parameters for the function $f_1(t)$ are
$b_1=0.42,\;\tau_1=8.8\times10^3,\;\tau_2=1.6\times10^5,\; 
\beta=0.73$ and for $f_2(t)$ are $c_1=0.14$ $a=5.6\times10^1$
$c_2=0.58$ and $\gamma=1.3\times10^3$.
 
Finally, note that the stretched exponential fit of relaxation
function  is more appropriate in this case due to the short
range of the potential.
We have found no evidence of logarithmic time dependence for
the energy relaxation in this case.
 
\section{Conclusion} 
 
Let us first summarize the main results of this work.
We have  demonstrated that the cooperative
dynamics of random walkers in a simple one-dimensional 
channel can give rise to highly non-exponential 
relaxation, when number density fluctuations are not
allowed.
In these simulations, two perturbing potentials (the Coulomb 
and the Lennard-Jones) have been used to study the effects of 
perturbing potential on the relaxation process.
The energy relaxation under these potentials can
be approximately described by a stretched exponential (in general)
in a closed system.
The variation in the time scale of relaxation  under 
these two well-known potentials  can be understood in terms
of the difference in the {\it range} of these potentials.
The Coulomb  potential being  long ranged (in comparison
with the Lennard-Jones interaction), shows much stronger
non-exponentiality in the energy relaxation.
Under the Coulomb potential, the exponent $\beta$ of
non-equilibrium energy relaxation function is 
0.52 while under Lennard-Jones potential it is 0.73.
The simulations seem to agree with the results of Bruns 
{\it et al.}\cite{berg} in showing a logarithmic time dependence
of the energy relaxation function under the Coulomb potential. 
However, under the short-ranged Lennard Jones potential,
the energy relaxation function does not show logarithmic
time dependence.

In the smaller sized systems the relaxation is found to
be faster and as the system size increases, relaxation 
slows down, as expected.
Also as expected, the density fluctuations in this
one-dimensional channel make the relaxation function 
faster and exponential.
When number fluctuation is allowed, random walkers overcome
the  resistance of the hard rod interactions, by moving
in and out of the linear lattice, such that the random
walkers experience no major hindrance to their flow.
In this case the system behaves like a system of weakly
interacting particles, without the need for strong
cooperativity for relaxation.
It is worth noting that interactions found in nature are
numerous, but the models of asymptotic time dependence of
relaxation commonly found in nature are limited in number.
However, the parameters of the relaxation function 
depend on the form of the biasing potential  and the 
nature of interaction between the carriers.

\section{Acknowledgment}

This work was supported in parts by grants from the Department of 
Atomic Energy (DAE) and the Council of Scientific and Industrial 
Research (CSIR), India.


\begin{references} 
\bibitem{berg}E. B. Bruns, M. L. Madaras, R. S. Coleman, 
C. J. Murphy and M. A. Berg, Phys Rev. Letts. 
{\bf 88} 158101 (2002). 
\bibitem{c1}S. Chandrasekhar Rev. Mod. Phys. {\bf 15}, 1 (1943) 
\bibitem{zew}S. K. Pal, L. Zhao, T. Xia and A. H. Zewail,
Proc. Natl. Acad. Sci. USA, {\bf 100} 13746 (2003).
\bibitem{c2}Non-equilibrium statistical mechanics in 
one dimension. Edited by V. Privman ( Cambridge University Press, 
Cambridge, 1997). 
\bibitem{c3}S. Alexander, J. Bernasconi, W. R. Schneider 
and R. Orbach Rev. Mod. Phys. {\bf 53}, 175 (1981).
\bibitem{c4}E. Barkai, V. Fleurov and J. Klafter 
Phys. Rev. E {\bf 61} 1164 (2000). 
\bibitem{hub1}D. L. Huber Phys. Rev. B. {\bf 15}, 533 (1977). 
\bibitem{rich1}P. M. Richards Phys. Rev. B. {\bf 16}, 1393 (1977).
\bibitem{rich2}P. M. Richards, R. L. Renken Phys. Rev. B  
{\bf 21 }, 3740  (1980). 
\bibitem{yu}K. W. Yu and P. M. Hui, Phys. Rev. A. 
{\bf 33}, 2745 (1986). 
\bibitem{pit}R. Pitis, Phys. Rev. B {\bf 48}, 4196 (1993).
\bibitem{kl1}G. Zumofen, J. Klafter, Phys. Rev. E {\bf 51},
2805 (1995).
\bibitem{kl2}A. Bar-Haim and J. Klafter J. Chem. Phys. {\bf 109}, 
5187 (1998). 
\bibitem{file1}C. Rodenbeck, J. Karger and K. Hahn, 
Phys. Rev. E {\bf 55} 5697 (1997).
\bibitem{file2}P. H. Nelson and S. M. Aurbach, J. Chem. Phys. 
{\bf 110} 9235 (1999).
\bibitem{file3}M. S.Okino, R. Q. Snurr, H. H. Kung, J. E. Ochs
 and M. L. Mavrovouniotis, J. Chem. Phys.{\bf 111} 2210 (1999).
\bibitem{kohl}R. Klohlrausch, Ann. Phys. (Leipzig) 
{\bf 12}, 393 (1847). 
\bibitem{fr}D. Frenkel, B. Smith, {\it Understanding molecular 
simulation: From algorithms to applications} (Academic Press,  
San Diego, 1996). 
\bibitem{ww} G. Williams and D. C. Watts, Trans. Faraday. Soc.
{\bf 66} 80 (1970).
\bibitem{got1}W. Gotze in Liquid Freezing and Glass Transition,
 Edited by D. Levesque, J. P. Hansen and J. Zinn-Justin
 (North Holland, Amsterdam, 1990). 
\bibitem{got2}W. Gotze and L. Sjogren Rep. Prog. Phys.{\bf 55},
241 (1992).
\bibitem{vlad}M. O. Vlad and M. C. Mackey J. Math. Phys. 
{\bf 36}, 1834(1995).
\bibitem{coh}M. W. Cohen and G. S. Grest, Phys. Rev. B, 
{\bf 24}, 4091 (1981). 
\bibitem{hub2}D. L. Huber, D. S. Hamilton and B. Barnett,  
Phys. Rev. B. {\bf 16}, 4642 (1977). 
\bibitem{hub3}D. L. Huber, Phys. Rev. B 
{\bf 31}, 6070 (1985),{\it ibid.} {\bf 53}, 6544 (1996). 
\end{references}
\end{document}